\documentclass[pre,twocolumn,showpacs]{revtex4}

\usepackage{graphicx}%
\usepackage{dcolumn}
\usepackage{amsmath}

\begin{document}

\title{Comment on ``Spin-1 aggregation model in one dimension''}

\author{Daniel Duque}%
 \email{campayo@phys.washington.edu}
\affiliation{%
Department of Physics\\
University of Washington\\
Seattle, WA 98195
}%

\date{\today}

\begin{abstract}
M. Girardi and W. Figueiredo have proposed a simple model
of aggregation in one dimension to mimic the self-assembly
of amphiphiles in aqueous solution [Phys. Rev. E \textbf{62},
8344 (2000)]. We point out that
interesting results can be obtained if
a different set of interactions is considered, instead of
their choice (the $s=1$ Ising model).
\end{abstract}

\pacs{82.60.Lf,64.75.+g,64.00.Ht}

\maketitle

M. Girardi and W. Figueiredo \cite{GF} (henceforth GF)
have considered the
$s=1$ Ising model in one dimension as a system
that could possibly show some of the features of micellar systems,
specially the existence of a critical micelle concentration (CMC).
Several definitions of the CMC can be provided, since this phenomenon
is actually a crossover, not a well defined phase transition.
Following the classic work of H. Wennerstr\"om and B. Lindman \cite{WL},
they focus on the distribution of aggregates with $n$ molecules,
$\rho(n)$. In real systems the curve $n\rho(n)$ is expected to show a
sharp maximum at $n=1$, an intermediate minimum and another maximum
at some $\bar{n}$ corresponding to the micelles, typically of the order
$\bar{n}\approx 10^2$.

This is indeed the case, at least in some regimes,
in the two- and three-dimensional models
previously studied by the same group \cite{GF2,GF3},
of which GF is a logical extension.
On those works the difference between
the minimum and the maximum at $\bar{n}$
was used as a sort of ``order parameter'', whose vanishing
could be associated with the CMC.
In the case studied in GF, it was found that this function is never
bimodal and so a CMC cannot be properly defined.

While their methods are not to be questioned, since they
have carried out Monte Carlo simulations and, most importantly,
it is possible to obtain an exact solution to the model,
we feel that the model considered
is not the best one if one wants to compare
with real amphiphilic systems and that the failure to
find a clear CMC seems natural in retrospect. As discussed in \cite{DT},
the mean feature of a micellar aggregate is the existence
of a preferred aggregate size, which in real systems is a
consequence of the effective geometry of the molecules
(as exemplified by the well know picture of Israelachvili's \cite{I}).
A model, even a highly simplified one,
should include some kind of preferred aggregate size in order
to present a CMC.

In a nutshell, the interaction considered in GF can
be described by the following choice of spin interactions:

\begin{center}
\begin{tabular}{c|ccc}
 & $\uparrow$  & $\downarrow$ & $\circ$   \\ \hline
 $\uparrow$    &    $ -J$    &   $J$ &$0$\\
  $\downarrow$ &    $J$      &  $-J$ &$0$\\
$\circ$        &    $0$      & $0$   &$0$,
\end{tabular}
\end{center}
where the arrows represent the $+1$ and $-1$ spins and the
circle the $0$ one.
Consider, instead, the following choice, in which we
choose a left-right picture of the spins, instead of
the usual up-down:
\begin{center}
\begin{tabular}{c|ccc}
 & $\leftarrow$  & $\rightarrow$ & $\circ$   \\ \hline
 $\leftarrow$    &    $0$    &  $-J$ &  $0$\\
  $\rightarrow$ &    $0$    &  $0$  & $-J$\\
$\circ$        &    $-J$   &  $0$  &  $0$.
\end{tabular}
\end{center}
We can expect this choice (which is basically an
application of the potential used in \cite{DT} to
a lattice) to provide
``micelles'' consisting of spin pairs with
orientation $\leftarrow\rightarrow$ on a ``solvent''
of $\circ$ spins, since this is the configuration
energetically favored. This is indeed the case:
\textit{e.g.}, we have checked that the slope of
the chemical potential as a function of the logarithm of
the amphiphilic density changes from $1$ to a value $1/2$,
and this crossover corresponds to the CMC \cite{DT};
this change in slope is not found with the GF model.

However, this model is too simple
to address the behavior of $n\rho(n)$, since there
is no possible aggregate between $n=1$ and $n=2$ where the
distribution would have a minimum. Our
model can then be trivially extended to another one with $m$ spins
($m-1$ orientations for the amphiphiles and one for the solvent)%
\footnote{%
We have also checked that extensions of the model in GF to
higher spin Ising and Potts models fail to produce CMC
features.}:
\begin{center}
\begin{tabular}{c|ccccc}
             & $\leftarrow$ &$\nwarrow$&$\cdots$&$\rightarrow$& $\circ$\\
\hline
$\leftarrow$ &    $0$       & $-J$     &$\cdots$& $0$         &  $0$\\
$\vdots$       &$\vdots$    & $\vdots$&$\ddots$   &$\vdots$         &$\vdots$ \\
$\rightarrow$&    $0$       & $0$&$\cdots$  & $0$           & $-J$\\
$\circ$      &    $-J$      & $0$&$\cdots$  &  $0$          &  $0$.
\end{tabular}
\end{center}

In Figure~1 we show the chemical potential \textit{versus} the logarithm
of the amphiphile density for the particular choice $m=5$ to show the
expected change in slope for $1$ to $1/4$. We also plot
include the corresponding values of $n\rho(n)$.
It is clearly seen how the feature sought in GF is indeed obtained
at $\ln(\rho)\approx -18$: at this point a unimodal distribution
with a maximum at $n=1$ turns into a bimodal with a second
maximum at $\bar{n}=4$ (in general, one finds $\bar{n}=m-1$);
the order parameter defined in GF is seen to vanish linearly.

\begin{figure}
\includegraphics[width=0.5\textwidth]{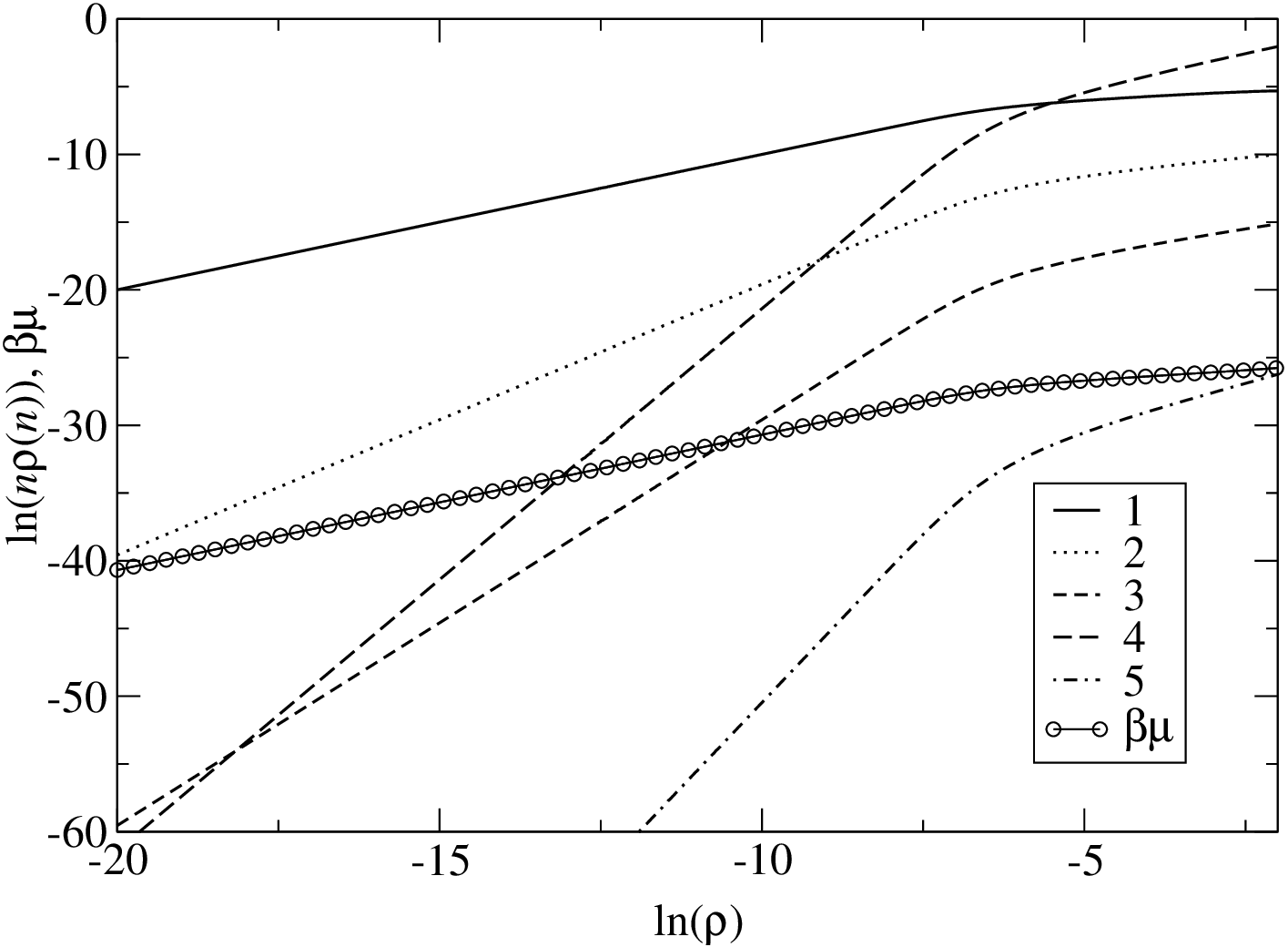}
\caption{%
Concentration of sites belonging to
aggregates with different $n$
(given in the legend) as
a function of amphiphilic density, given as
$\ln(n\rho(n))$
\textit{vs.} $\ln(\rho)$, we also plot the
chemical potential, $\beta\mu$;
$m=5$ and $\beta J=20$.}
\end{figure}

However, this point is quite
far from the CMC one would define from the chemical
potential,
$\ln(\rho)\approx -6$; on the other hand, the
crossing between the $n=1$ and the $\bar{n}=4$ lines
does lie in this range.
We suggest that this kind of criterion for the CMC
(in general, $\bar{n}\rho(\bar{n})=\rho(1)$, 
so that
an order parameter could be defined as
the difference between the $n=1$ and
the $\bar{n}$ maxima) should perhaps be a better choice that the
one in GF and previous works (Refs. \cite{GF2,GF3}).

\begin{figure}
\includegraphics[width=0.5\textwidth]{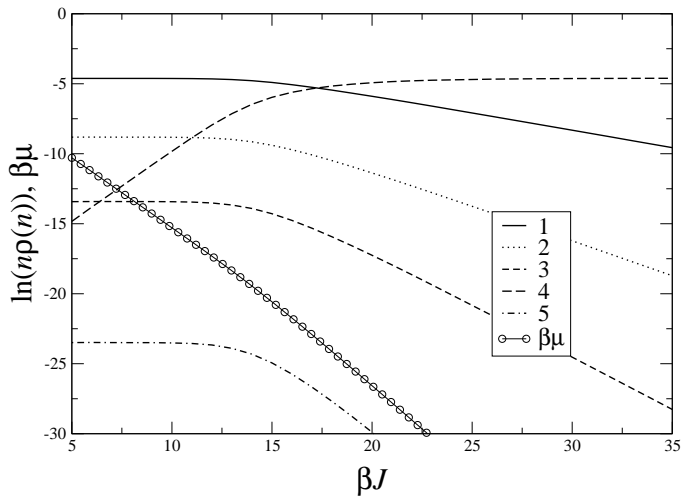}
\caption{%
Same as Fig.~1 but now
for a fixed amphiphilic density $\rho=0.01$
and varying $\beta J$.}
\end{figure}

We can also fix the concentration at some value
and vary the interaction parameter $\beta J$. In Figure~2 we show
results for $\rho=0.01$. The results are quite similar to
the ones in Fig.~1, except that the chemical potential is
now determined by our choice of density and does not show
any particular feature. These two choices of crossing the
CMC line yield the same results, which we have combined
in Figure~3. There we present results for $m=5$ and also
for $m=10$. We can see the CMC lines with the
$\bar{n}\rho(\bar{n})=\rho(1)$ criterion show
the expected dependence on $m$, with the slopes approaching
a slope $m-2$ at high interactions and low densities (since
each aggregate contains $m-2$ favorable bonds). The
other criterion, based on the bimodality of the
distribution (\textit{i.e.},
$\bar{n}\rho(\bar{n})=(\bar{n}-1)\rho(\bar{n}-1)$)
is seen to provide very different results, with
a limiting slope of $1$, independent on $m$.
Nevertheless, the discrepancies between
these two criteria can be expected to be smaller
in real amphiphilic systems, since the
energy will quickly grow for aggregates
either smaller or larger than $\bar{n}$, not
linearly (for $n<\bar{n}$) as in our case.

We kindly acknowledge Prof. Pedro Tarazona's
helpful suggestions. This research
has been supported by a Spanish Ministry of Education, Culture and
Sport MEC-FPI EX 2000 grant.


\begin{figure}
\includegraphics[width=0.5\textwidth]{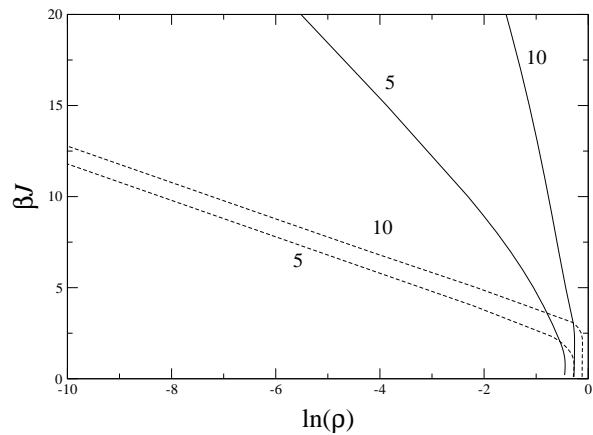}
\caption{%
Location of the CMC in the $\ln(\rho)$--$\beta J$
plane, defined as
$\rho(1)=\bar{n}\rho(\bar{n})$ 
(solid lines) and as
$\bar{n}\rho(\bar{n})=(\bar{n}-1)\rho(\bar{n}-1)$ 
(dashed lines), both for $m=5$ and $m=10$.}
\end{figure}


\begin{thebibliography}{6}
\expandafter\ifx\csname natexlab\endcsname\relax\def\natexlab#1{#1}\fi
\expandafter\ifx\csname bibnamefont\endcsname\relax
  \def\bibnamefont#1{#1}\fi
\expandafter\ifx\csname bibfnamefont\endcsname\relax
  \def\bibfnamefont#1{#1}\fi
\expandafter\ifx\csname citenamefont\endcsname\relax
  \def\citenamefont#1{#1}\fi
\expandafter\ifx\csname url\endcsname\relax
  \def\url#1{\texttt{#1}}\fi
\expandafter\ifx\csname urlprefix\endcsname\relax\def\urlprefix{URL }\fi
\providecommand{\bibinfo}[2]{#2}
\providecommand{\eprint}[2][]{\url{#2}}

\bibitem[{\citenamefont{Girardi and Figueiredo}(2000{\natexlab{a}})}]{GF}
\bibinfo{author}{\bibfnamefont{M.}~\bibnamefont{Girardi}} \bibnamefont{and}
  \bibinfo{author}{\bibfnamefont{W.}~\bibnamefont{Figueiredo}},
  \bibinfo{journal}{Phys. Rev.\ E} \textbf{\bibinfo{volume}{62}},
  \bibinfo{pages}{8344} (\bibinfo{year}{2000}{\natexlab{a}}).

\bibitem[{\citenamefont{Wennerstr{\"o}m and Lindman}(1979)}]{WL}
\bibinfo{author}{\bibfnamefont{H.}~\bibnamefont{Wennerstr{\"o}m}}
  \bibnamefont{and} \bibinfo{author}{\bibfnamefont{B.}~\bibnamefont{Lindman}},
  \bibinfo{journal}{Phys. Rep.} \textbf{\bibinfo{volume}{52}},
  \bibinfo{pages}{1} (\bibinfo{year}{1979}).

\bibitem[{\citenamefont{de~Moraes and Figueiredo}(1999)}]{GF2}
\bibinfo{author}{\bibfnamefont{J.}~\bibnamefont{de~Moraes}} \bibnamefont{and}
  \bibinfo{author}{\bibfnamefont{W.}~\bibnamefont{Figueiredo}},
  \bibinfo{journal}{J. Chem. Phys.} \textbf{\bibinfo{volume}{110}},
  \bibinfo{pages}{2264} (\bibinfo{year}{1999}).

\bibitem[{\citenamefont{Girardi and Figueiredo}(2000{\natexlab{b}})}]{GF3}
\bibinfo{author}{\bibfnamefont{M.}~\bibnamefont{Girardi}} \bibnamefont{and}
  \bibinfo{author}{\bibfnamefont{W.}~\bibnamefont{Figueiredo}},
  \bibinfo{journal}{J. Chem. Phys.} \textbf{\bibinfo{volume}{112}},
  \bibinfo{pages}{4833} (\bibinfo{year}{2000}{\natexlab{b}}).

\bibitem[{\citenamefont{Duque and Tarazona}(1997)}]{DT}
\bibinfo{author}{\bibfnamefont{D.}~\bibnamefont{Duque}} \bibnamefont{and}
  \bibinfo{author}{\bibfnamefont{P.}~\bibnamefont{Tarazona}},
  \bibinfo{journal}{J. Chem. Phys.} \textbf{\bibinfo{volume}{107}},
  \bibinfo{pages}{10207} (\bibinfo{year}{1997}).

\bibitem[{\citenamefont{Israelachvili}(1985)}]{I}
\bibinfo{author}{\bibfnamefont{J.}~\bibnamefont{Israelachvili}}, in
  \emph{\bibinfo{booktitle}{Physics of Amphiphiles: Micelles, Vesicles and
  Microemulsions}}, edited by
  \bibinfo{editor}{\bibfnamefont{V.}~\bibnamefont{Degiorgio}} \bibnamefont{and}
  \bibinfo{editor}{\bibfnamefont{M.}~\bibnamefont{Corti}}
  (\bibinfo{publisher}{North-Holland}, \bibinfo{year}{1985}).

\end{thebibliography}
\end{document}